\documentclass[a4paper]{jpconf}

\usepackage{epsfig,amsmath}
\newcommand{\rd}{\rho_D^3}

\newcommand{\as}{\alpha_s}

\newcommand{\qqh}{\hat{q}^2}

\newcommand{\qq}{q^2}
\newcommand{\qz}{q_0}
\newcommand{\el}{E_\ell}

\newcommand{\GeV}{\,\mbox{GeV}}

\def \be{\begin{equation}}
\def \ee{\end{equation}}
\newcommand{\bea}{\begin{eqnarray}}
\newcommand{\eea}{\end{eqnarray}}

\begin{document}
%\begin{titlepage}

\title{Semileptonic $B$ decays
and the inclusive determination of $|V_{ub}|$}

\author{Paolo Giordano}
\address{Dip.\ Fisica Teorica, Univ.\ di Torino, \& INFN  Torino,
I-10125 Torino, Italy}
\ead{pgiordan@to.infn.it}

\begin{abstract}
We present a new theoretical framework for the study of
$B\to X_u \ell \nu$ decays, which includes all known
perturbative and non-perturbative contributions and
a description of leading and subleading Fermi motion effects.
The perturbative and non-perturbative regimes are
separated by a ``hard'' Wilsonian cutoff $\mu \sim 1$ GeV.
We bring into focus some problems related to the high $q^2$ region
and to Weak Annihilation effects. We provide estimates of the
CKM parameter $|V_{ub}|$ using the described framework and
discuss the related theoretical uncertainty.
\end{abstract}

%----------------------------------------------------------------------------

\section{Introduction}
The precise determination of the element $|V_{ub}|$ of the CKM matrix is an
important test of the flavour structure of the Standard Model (SM) and is crucial
in the indirect search for New Physics. 
The latest global fit to the Unitarity Triangle (UT) including 
all flavour changing observables but a direct
determination of $|V_{ub}|$ predicts
$|V_{ub}|= (3.44\pm0.16)\times 10^{-3}$ \cite{UTfit}. This value agrees within errors
with the {\it exclusive} determination, that relies on lattice QCD or
light-cone sum rules \cite{Flynn,Ball} and that is still affected by somewhat large 
theoretical errors. An {\it inclusive} analysis is in principle the cleanest method
to precisely determine $|V_{ub}|$. This is based on the comparison between the
decay rate of $B\to X_u \ell \nu$ measured by experiments and the
corresponding theoretical prediction.
The latest HFAG world average \cite{HFAG} yields an inclusive 
$|V_{ub}|$ which is about $2.5 \sigma$ higher than the value preferred by the
global UT fit, calling for a deeper investigation of the process.

The theoretical description of $B\to X_u \ell \nu$ is based on a local
Operator Product Expansion. Inclusive quantities are organized in a 
double series in $\as$ (perturbative QCD corrections) 
and in $1/m_b$ (Heavy Quark Expansion).
The very same method was successfully applied to the $b \rightarrow c$ decay 
and led to a precise determination of $|V_{cb}|$, within 2\%. The description of
charmless decays is more involved due to the dominant charmed background that
needs to be rejected by experiments imposing very stringent cuts. These cuts
can spoil the convergence of the OPE introducing sensitivity to nonlocal
effects, such as the motion of the $b$ quark inside the meson (Fermi motion),
that can be parameterized in terms of a light-cone distribution function 
(or ``shape function''). The
lowest integer moments of the distribution function
are constrained by the OPE \cite{Bigi:1993ex} and they are expressed
in terms of the $b$ quark mass and of the same 5 and 6 dimensional operators
that contribute to $B\to X_c \ell \nu$. Such expressions are universal, 
i.e. independent of the process, and shared by the
radiative decay  $B\to X_s \gamma$ only as long as $1/m_b$ corrections are neglected.  

An OPE-based treatment of shape function effects in $B\to X_s \gamma$ 
including subleading ($1/m_b$) effects was developed
in \cite{benson} and turned out to describe well experimental data. 
In \cite{Gambino:2007rp} a similar procedure was undertaken for the case 
of semileptonic decays, where many complications arise, mostly due to the kinematics
taking place at different $q^2$. In the
following we illustrate the main features of this procedure and show some 
meaningful results. 

\section{Theoretical framework}
\subsection{Perturbative corrections in a Wilsonian approach}
All observables describing the $B\to X_u \ell \nu$ decay can be
extracted {\it via} integration over the triple differential width:
\begin{eqnarray} \label{eq:aquila_normalization}
\frac{d^3 \Gamma}{d\qq \,d\qz \,d\el}  &=&
\frac{G_F^2 |V_{ub}|^2}{8\pi^3}
 \Bigl\{ 
\qq W_1- \left[ 2\el^2-2\qz \el + \frac{\qq}{2} \right] W_2 \nonumber 
+  \qq (2\el-\qz) W_3 \Bigr\}\times \\&&
\quad\quad\quad\quad    \times \theta \left(\qz-\el-\frac{\qq}{4\el} \right) 
 \ \theta(\el) \ \theta(\qq) \ \theta(\qz-\sqrt{\qq}),
\end{eqnarray} 
where $q_0$ and $E_\ell$ are the total leptonic and the charged lepton energies
in the $B$ meson rest frame, $q^2$ is the leptonic invariant mass and $W_{1-3}$
are the three structure functions relevant in the case of massless lepton.

Perturbative corrections to the structure functions $W_{1-3}$ to order ${\cal O}(\as)$ have
been known for quite long \cite{dfn}, whereas ${\cal O}(\as^2 \beta_0)$ corrections
recently appeared in \cite{Gambino:2006wk}. Both calculations were performed in the
{\it on-shell} scheme.

It has been stressed several times in the literature \cite{kinetic} that an
{\it on-shell} definition of the $b$ quark mass becomes ambiguous as soon as
power suppressed terms are included. The {\it pole} mass is better traded with
a {\it running} mass $m_b(\mu)$. To this purpose, perturbative corrections to
order ${\cal O}(\as^2 \beta_0)$ are calculated anew in \cite{Gambino:2007rp}
in the presence of a ``hard''  Wilsonian cutoff $\mu \sim 1$ GeV. This
new scale separates the perturbative regime of gluons with energies larger than $\mu$  
from the ``soft'' (non-perturbative) regime of gluons with energies lower than $\mu$. 
The contributions of soft gluons are then
absorbed into a redefinition of the heavy quark parameters, consistent with the way
they are extracted from fits to $B\to X_c \ell \nu$ moments in the
{\it kinetic} scheme \cite{Gambino:2004qm,BF}. Physical observables are of course
independent of the cutoff.

\subsection{Fermi motion}
As already mentioned in the Introduction, Fermi motion is encoded in a distribution
function, whose lowest integer moments are constrained by the local OPE.
As soon as $1/m_b$ corrections are retained, such moments cease to be universal:
they have different expressions for each of the three structure functions in eq.
\eqref{eq:aquila_normalization} and show an explicit $q^2$ dependence. To preserve
generality we introduce three separate distribution functions, one for each
of the structure functions, depending on the light-cone component of the 
$b$ quark momentum ($k_+$), on $q^2$ and on the Wilsonian cutoff ($\mu$). Hadronic structure
functions are then defined {\it via} a convolution of the perturbative structure functions
with the distribution functions, whose expression is derived at the leading order in $1/m_b$
and $\as$ and assumed to be valid also at higher orders:
\begin{equation} \label{eq:conv2}
 W_i(q_0,q^2) \sim
 \int dk_+ \ F_i(k_+,q^2,\mu) \ W_i^{pert}
\left[ q_0 - \frac{k_+}{2} \left( 1 - \frac{q^2}{m_b M_B} \right), q^2,\mu \right]
\end{equation}
Model-dependence resides only in the {\it Ansatz} employed for
the distribution functions. In the analysis of \cite{Gambino:2007rp} a set of
about 80 different functional forms, inspired by those already present
in the literature (exponential, Gaussian, Roman, hyperbolic), is tested and the
related uncertainty on $|V_{ub}|$ turns out to be rather small 
(see Sec. \ref{results}). 

\subsection{The high $q^2$ region}
The impact of Fermi motion becomes irrelevant at high $q^2$ and the developed
formalism is no more applicable. The OPE itself shows a number of pathological 
features in this kinematical region, due to the emergence of unsuppressed higher order
terms. For instance, the OPE predicts a value for the variance of
the distribution functions which decreases at increasing $q^2$, reaching
even negative values. Moreover, Wilson coefficients
of power suppressed operators become more and more important
and already the coefficient of the Darwin term ($\rd$) shows a divergence at $q^2 = m_b^2$:
\be
\frac{d\Gamma}{d\qqh}\sim  \frac{\rd}{6m_b^3}\left[
20\, {\hat q}^6 +66\, {\hat q}^4+48\, \qqh+74 -\frac{96}{1-\qqh}\right]+...,
\quad \qqh=\frac{q^2}{m_b^2}
\label{q2ope}
\ee 
This singularity is removed at the level of the total rate 
by a one-loop penguin diagram that mixes
the Weak Annihilation (WA) four-quark operator into the Darwin operator 
\cite{bumoments,Ossola:2006uz}. However, as we are interested in differential distributions
as well, a dedicated treatment of the high-$q^2$ region ($q^2 > q^2_* \sim 11\ \text{GeV}^2 $) is
mandatory. 
In \cite{Gambino:2007rp} two different methods are proposed and their difference
is used to estimate the associated uncertainty: 
\begin{itemize}
\item[{\it a)}] we model the tail in a way consistent
with positivity of the spectra and including a WA contribution ($X$)
through a Dirac-$\delta$ localized at the endpoint (default method):
\be \frac{d\Gamma}{d\qqh}\sim
\frac{\rd}{6m_b^3}\left[ 20\, {\hat q}^6 +66\, {\hat q}^4+48\, \qqh+74
-\frac{96\,(1-e^{-\frac{(1-\qqh)^2}{b^2}} )} {1-\qqh}\right]+ X\,
\delta(1-\qqh)+...
\label{q2mod}
\ee 
\item[{\it b)}] we extend the Fermi motion description of low $q^2$ to the
high $q^2$ region, {\it freezing} the shape function at $q^2 = {q_*}^2$
and using it in the convolution of eq. \eqref{eq:conv2} at higher $q^2$. 
\end{itemize}

\section{Results and theoretical uncertainties} \label{results}
We take advantage of some of the latest experimental
measurements to extract values of $|V_{ub}|$ using the descripted
framework. We leave the task of an average of these results to a future,
hopefully dedicated, experimental analysis. We consider:
\begin{itemize}
\item [\bf A] Belle analysis with $M_X\le1.7\GeV$ and $E_\ell>1.0 \GeV$ \cite{belle1}; 
\item  [\bf B] Belle and Babar analyses with  $M_X\le1.7\GeV$, $q^2>8\GeV^2$, 
and $E_\ell>1.0 \GeV$ \cite{belle1,babar2};
\item [\bf C] Babar analysis with $E_\ell>2.0\GeV$ \cite{babar1} 
\end{itemize}
\begin{table}[t]
\center{
\begin{tabular}{|c|c|c|c|c|c|c|c|c|c|c|}
\hline
cuts & {\small $|V_{ub}|\times 10^3$} & $f$  & exp & par &  pert & {\small
tail model}
& $q_*^2$ & $X$ & ff &tot th\\ \hline
{\bf A}  \cite{belle1} & 3.87 & 0.71 & 6.7  & 3.5 & 1.7 & 1.6 &
2.0  &$_{-2.7}^{+0.0}$ &$^{+2.4}_{-1.1}$ & $\pm 4.7^{+2.4}_{-3.8}$   \\
\hline
{\bf B}  \cite{belle1,babar2} & 4.44 & 0.38 & 7.3 & 3.5 & 2.6 & 3.0 & 4.0
& $_{-5.0}^{+0.0}$ & $^{+1.4}_{-0.5}$ & $\pm 6.6_{-5.5}^{+1.4}$  \\ \hline
{\bf C}  \cite{babar1} & 4.05 & 0.30 & 5.7 & 4.2 &3.3  & 1.8 & 0.9
&$_{-6.2}^{+0.0}$  &  $^{+1.2}_{-0.7}$ & $\pm 5.7^{+1.2}_{-6.9}$  \\
\hline
\end{tabular}}
\caption{\small Values of $|V_{ub}|$ obtained using different experimental results
 and their experimental and theoretical 
uncertainties (in percentage) due to various sources (see text). 
$f$ is the estimated fraction of events. }
\end{table}
Results are summarized in Table 1 and were obtained from a C++ code
available upon request. The reported values of $|V_{ub}|$
are obtained using the default setting, namely an exponential {\it Ansatz}
for the distribution functions, the prescription {\it a)} for the high $q^2$ tail
at $X=0$
(see previous section) and the central values of the fit in \cite{BF} as input parameters, at
$\mu$ = 1 GeV. $f$ is the estimated fraction of events, whereas the
following columns show different sources of uncertainty, namely: 
\begin{itemize}
\item Experimental error (exp).
\item Parameteric error (par): it is extracted taking into account all correlations
between non-perturbative parameters \cite{BF} and varying
$\alpha_s = 0.22$ by $\pm 0.02$ as uncorrelated. The uncertainty on $m_b$ is
by far dominating.
\item Perturbative error (pert).
\item Errors related to the high $q^2$ region: we consider the difference between
methods {\it a)} and {\it b)} (tail model), the value of $q^2$ at which the
modelling sets in ($q_*^2$) and the error due to WA effects ($X$).
We let $X$ vary in a range consistent with
the 90\% confidence level bound set by CLEO on the size of WA \cite{cleo_WA},
namely $0 \le X \le 0.04$. We stress that the error related to $X$ is asymmetric and
points to a lower value of $|V_{ub}|$.
\item Functional form dependence (ff), estimated using about 80 different
{\it Ans$\ddot a$tze} for the distribution functions.
\end{itemize}
It is worth stressing that all combinations of cuts considered include the
high $q^2$ region discussed in the previous section which, as we showed, 
is plagued by poorly controlled effects. However, Belle measurements {\bf A} and
{\bf B} can be easily combined in order to obtain an estimate of $|V_{ub}|$
with an {\it upper} cut on $q^2$, namely for the combination $M_X\le1.7\GeV$,
$E_\ell>1.0 \GeV$, and $q^2<8\GeV^2$. This yields, within our framework, a value 
of $|V_{ub}|$ much lower than in all other cases ($|V_{ub}|=3.18\times 10^{-3}$).
and it might signal some bias in the treatment of the high $q^2$ region either
on the experimental or on the theoretical side. Only a dedicated experimental analysis
with an upper cut on $q^2$ could probably shed some light on this issue.

\section{Summary and References}
We presented a new approach for dealing with the triple differential width of 
$B \to X_u \ell \nu$ decays, in a framework characterised by a hard Wilsonian cutoff 
$\mu \sim 1$ GeV. The method developed takes into account all known perturbative and
non-perturbative corrections. Fermi motion is treated at the subleading level as
well.
Some problems related to the high $q^2$ region of the process were pointed out, that
were probably underestimated in the past and that still deserve a deeper
investigation. 
We also presented some numerical results with the associated uncertainties and
put forward the suggestion of a new experimental analysis with an upper cut
on $q^2$. 

%\section*{References}

\end{document}